\documentclass[11pt,a4paper,reqno]{amsart}

\usepackage[english]{babel}
\usepackage{siunitx}
\usepackage{cite}
\usepackage{upgreek}
\usepackage{tabularx}
\usepackage{graphicx} 
\usepackage[headings]{fullpage}
\usepackage{amsaddr}
\usepackage{dsfont}
\usepackage{xcolor}
\usepackage[singlelinecheck=false]{caption}
\usepackage{tikz}
\usetikzlibrary{arrows.meta}
\usepackage{url}

\newcolumntype{Y}{>{\centering\arraybackslash}X}

\newcommand{\R}{\mathbb{R}}
\newcommand{\N}{\mathbb{N}}

\DeclareSIUnit\bar{bar}
\newcommand\Tstrut{\rule{0pt}{2.6ex}}         
\newcommand\Bstrut{\rule[-0.9ex]{0pt}{0pt}}

\begin{document}
\title[]{Data-driven stochastic 3D modeling of the nanoporous binder-conductive additive phase in battery cathodes$^\star$}

\author{Phillip~Gräfensteiner$^{1}$, Markus~Osenberg$^2$, Andr\'{e}~Hilger$^2$, Nicole~Bohn$^3$, Joachim R. Binder$^3$, Ingo~Manke$^2$, Volker~Schmidt$^1$,~Matthias~Neumann$^{4}$}

\address{$^1$Institute of Stochastics, Ulm University, 89069~Ulm, Germany\footnote{$^\star$ This preprint has not undergone any post-submission improvements or corrections. The Version of Record of this article is published in \emph{Journal of Mathematics in Industry}, and is available online at \url{https://doi.org/10.1186/s13362-025-00174-z}}}
\address{$^2$Institute of Applied Materials, Helmholtz-Zentrum Berlin f\"ur Materialien und Energie, Hahn-Meitner-Platz~1, 14109 Berlin, Germany}
\address{$^3$Institute for Applied Materials, Karlsruhe Institute of Technology, Hermann-von-Helmholtz-Platz 1, 76344 Eggenstein-Leopoldshafen, Germany}
\address{$^4$Institute of Statistics, Graz University of Technology, Kopernikusgasse 24/III, \\ 8010 Graz, Austria}
\email{neumann@tugraz.at (corresponding author)}



\begin{abstract}
A stochastic 3D modeling approach for the nanoporous binder-conductive additive phase in hierarchically structured cathodes of lithium-ion batteries is presented. The  binder-conductive additive  phase of these electrodes consists of carbon black, polyvinylidene difluoride binder and graphite particles. For its stochastic 3D modeling, a three-step procedure based on methods from stochastic geometry is used. First, the graphite particles are described by a Boolean model with ellipsoidal grains. Second, the mixture of carbon black and binder is modeled by an excursion set of a Gaussian random field in the complement of the graphite particles. Third, large pore regions within the mixture of carbon black and binder are described by a Boolean model with spherical grains. The model parameters are calibrated to 3D image data of cathodes in lithium-ion batteries acquired by focused ion beam scanning electron microscopy. Subsequently, model validation is performed by comparing model realizations with measured image data in terms of various morphological descriptors that are not used for model fitting. Finally, we use the stochastic 3D model for predictive simulations, where we generate virtual, yet realistic, image data of nanoporous binder-conductive additives with varying amounts of graphite particles. Based on these virtual nanostructures, we can investigate structure-property relationships. In particular, we quantitatively study the influence of graphite particles on effective transport properties in the nanoporous binder-conductive additive  phase, which have a crucial impact on electrochemical processes in the cathode and thus on the performance of battery cells.
\end{abstract}

\maketitle

\section{Introduction}
 Batteries play an essential role in the transition of the power sector to renewable energies~\cite{passerini.2020}. Thus, an improvement of battery performance is an important challenge. Battery performance depends on physical properties of the electrode materials, such as effective conductivity of electrons or effective ionic diffusivity. As for heterogeneous materials in general~\cite{torquato.2013}, these physical properties are strongly influenced by the morphology of micro- and nanostructures within electrode materials~\cite{jain.2022}. 
At this, a crucial factor with respect to cathode materials is--besides the spatial arrangement of active material--the morphology of conductive additives and binder~\cite{entwistle.2022,hein.2020a, prifling.2022}. The latter influences effective conductivity of electrons or effective ionic diffusivity in the cathode. Based on 3D image data, it is possible to quantitatively characterize the morphology of the binder-conductive additive phase, in the following referred to  only as additive phase, and link morphological descriptors to effective properties, which can be numerically simulated~\cite{kroll.2021,cadiou.2020a,cadiou.2020b,Mistry.2021,prifling.2022}. However, highly-resolved 3D imaging techniques are expensive in costs and time and thus, only a small number of samples can be studied in this way. 

One approach to overcome this problem is to model the 3D micro- and nanostructure of material samples, using tools of stochastic geometry and mathematical morphology~\cite{chiu.2013, jeulin.2021, lantuejoul.2002, schmidt.2015}. 
Doing so, one can generate so-called digital twins of micro- and nanostructures, which are statistically similar to those observed by 3D imaging.
In the present paper, we follow the approach considered in~\cite{Kalidindi.2022} in defining a digital twin as “a high-fidelity in-silico representation closely mirroring the form (\emph{i.e.}, appearance) and the functional response of a specified (unique) physical twin”. In this context, a physical twin is the physical sample of the micro- or nanostructure under consideration, experimentally measured by 3D imaging. Then, 
based on a physical twin, a parametric stochastic 3D model is calibrated, allowing for the generation of virtual structures that mirror the appearance and functional behavior of the physical twin. Finally, model validation is performed with respect to morphological descriptors that were not used for model fitting in order to ensure that the generated virtual structures can be considered as digital twins.


Previous studies involving virtual electrode structures can be broadly classified into two categories.
On the one hand, there are parametric approaches for generating virtual electrode structures including the additive phase of battery materials~\cite{Chen.2007,Srivastava.2020}, where the underlying models are not calibrated to experimental image data. On the other hand, there are non-parametric generative models based on machine learning that allow for the generation of virtual electrode structures~\cite{dahari.2023, gayon-lombardo.2020} that nicely reproduce image data, but lack interpretability and control over altering the resulting structures. In the present paper, we propose a combination of the above through a low-parametric data-driven modeling approach for the 3D morphology of the additive phase in electrodes of lithium-ion batteries, which consists of carbon black, polyvinylidene difluoride (PVDF) binder, graphite particles and pores. This low-parametric approach allows for a high interpretability of  model parameters. In particular, we can control the amount of graphite particles within the additive phase by just one model parameter.

For the 3D  modeling approach considered in the present paper, we combine two different  model types from stochastic geometry~\cite{chiu.2013}, namely excursion sets of random fields with so-called Boolean models. Note that excursion sets of random fields have been used to model various types of micro- and nanostructures in electrodes of solid oxide fuel cells~\cite{marmet.2023, moussaoui.2018, moussaoui.2019, abdallah.2015, n.2018b} and lithium-ion as well sodium-ion batteries~\cite{neumann.2023, neumann.2024, prifling.2021b}. Boolean models are also frequently used for generating virtual micro- or nanostructures such as, \emph{e.g.}, porous membranes~\cite{prill.2020, roldan.2021}, and rubber with carbon black~\cite{jean.2011a, jean.2011b}. 

\begin{figure}[h]
	\centering
	\includegraphics[width=0.75\textwidth]{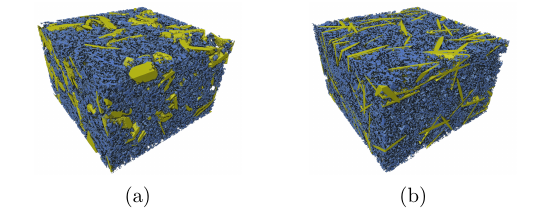}
	\caption{3D rendering of tomographic image data, \emph{i.e.}, a physical twin (a), and of a realization of the fitted stochastic 3D  model, \emph{i.e.}, a digital twin (b).
    Graphite particles are visualized in yellow, while carbon black and PVDF binder are depicted in blue. The physical size of the cutouts is $9\times 14\times 14$~\unit{\micro\metre\cubed}.}
	\label{fig:3Dvis}
\end{figure}

For the additive phase of the  hierarchically structured cathodes of lithium-ion batteries considered in the present paper, we propose a three-step approach. First, the graphite particles are described by a Boolean model with ellipsoidal grains. 
In a second step, the mixture of carbon black and binder is modeled as an excursion set of a Gaussian random field in the complement of the graphite particles. Third, large pore regions are described by a Boolean model with spherical grains. The model parameters are calibrated based on tomographic image data of cathodes in lithium-ion batteries acquired by focused ion beam scanning electron microscopy (FIB-SEM)~\cite{holzer.2012,moebus.2007}. For this, after a phase-based segmentation of the gray scale images, where each voxel is either assigned as pore or solid, the solid phase is further segmented by machine learning and thereby subdivided into graphite particles and a mixture of carbon black and binder. Then, the resulting three-phase image data is the basis for model calibration. 
By means of the calibrated stochastic 3D model, we are able to generate virtual, but realistic additive phases. A visual comparison between a model realization and  segmented image data is provided in Figure~\ref{fig:3Dvis}.

Model validation is performed by comparing realizations drawn from the calibrated model with tomographic image data in terms of various transport-relevant morphological descriptors that are not used for model calibration. In particular, we consider mean geodesic tortuosity and constrictivity. These two morphological  descriptors quantify the length of shortest paths through a given material phase as well as bottleneck effects. They have been shown to strongly influence effective transport properties of porous materials~\cite{ n.2019, prifling.2021c}.

Additionally, we  generate and validate model realizations at voxel resolutions that are coarser than the one of tomographic FIB-SEM image data, which is used for model calibration. 
This is motivated by the fact that many numerical simulation algorithms cannot handle the amount of detail in highly-resolved image data, and are bottlenecked by computational limitations. We therefore generate model realizations at a voxel resolution that is coarser by a factor of $2$ and $4$, respectively, where we study the effect of lowering the resolution in two different ways. First, we quantify the influence on the morphological descriptors considered in this paper and, second, we study whether the model fit is still good enough when comparing down-sampled model realizations with down-sampled image data.

Finally, we use the stochastic 3D  model to study structure-property relationships. Note that the low-parametric nature of our modeling approach allows us to control the generation of virtual nanostructures in an interpretable way. In particular, we are able to generate virtual additive phases with varying amounts of graphite. Furthermore, we perform a simulation study to investigate the impact of the amount of graphite on morphological descriptors and effective transport properties.

The rest of this paper is organized as follows. In Section~\ref{sec:materials}, a description of the  cathode material along with some details on image acquisition, data post-processing and segmentation are given. Then, in Section~\ref{sec:model_desc}, we describe the proposed stochastic 3D model in detail. The estimation procedure of the involved model parameters based on tomographic image data is presented in Section~\ref{sec:modelcalib}, whereas Section~\ref{sec:modelval}  deals with model validation. The framework of a simulation study for varying amounts of graphite is explained in Section~\ref{sec:vary_graphite}. A detailed discussion on the obtained results is given in Section~\ref{sec:discussion}. Finally, Section~\ref{sec:conclusion} concludes.

\section{Materials and 3D imaging}\label{sec:materials}
\subsection{Sample preparation}\label{sec:sample_preparation}
\begin{figure}[h]
    \centering
    \includegraphics[]{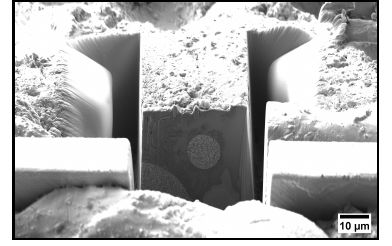}
    \caption{Visualization of sample prepared for FIB-SEM imaging.}
    \label{fig:sample_vis}
\end{figure}

The cathode investigated in the present paper consists of 87~wt\% active material, 5~wt\% graphite, 4~wt\% carbon black and 4~wt\% PVDF binder. Therefore, PVDF binder (Solef 5130, Solvay Solexis), carbon black (Super C65, Imerys Graphite\&Carbon), graphite (KS6L Imerys Graphite\&Carbon), and nano porous NCM powder were dispersed in N-methyl-2-pyrrolidone (Sigma Aldrich), where the NCM powder is manufactured as described in~\cite{wagner.2020}. 
The slurry was cast on a \SI{20}{\micro\metre} thick aluminum foil at \SI{350}{\micro\metre} gap height and dried overnight at \SI{80}{\celsius}. After drying, the thickness of the cathode was about \SI{120}{\micro\metre}, see Figure~\ref{fig:sample_vis}.

\subsection{3D imaging}\label{sec:imaging}
FIB-SEM imaging has been performed for the cathode sample manufactured as described in Section~\ref{sec:sample_preparation}. The sample exhibits an open, i.e. interconnected, pore space. In order to avoid shine-through artifacts in the resulting image data, the pore space was infiltrated with ELASTOSIL\textsuperscript{\tiny\textregistered} RT601 (WACKER) resin. The sample was infiltrated at \SI{200}{\milli\bar} low pressure and then cured at normal pressure and room temperature.
After 24 hours, the silicone-infiltrated specimen was embedded in EpoThin 2 resin to provide additional stiffness for the final grinding and polishing steps. 
Polishing of the sample led to a cross section of the cathode for further examination. 
First, the sample was fixed to a standard microscopy aluminum stub using a silver conductive adhesive and then coated in gold. The 3D tomography measurement was performed using a Zeiss crossbeam 340 focused ion beam (FIB) at the Corelab Correlative Spectroscopy and Microscopy Laboratory (CCMS, HZB, Germany). 
After transferring the prepared sample under dry air conditions, a U-shape as shown in Figure~\ref{fig:sample_vis} was milled into the exposed cross section using a \SI{30}{\nano\A} gallium current at \SI{30}{\kilo\eV}. 
During this process, a part of the aluminum conductor had to be removed by FIB milling as it blocked the view to the central part of the U-shape. This unmodified central part, around which the U-shape was milled, had a diameter of \SI{35}{\micro\metre}. The front of this part, which later formed the first slice of the tomography, was then polished using a \SI{1.5}{\nano\A} gallium current. 
The pixel size of 2D images obtained by the SEM, operating at \SI{1}{\kilo\eV}, was set to $(\SI{10}{\nano\metre})^2$ and the dynamic focus and tilt compensation were adjusted. To reduce the measurement time, the slice thickness of the serial section for tomography was set to \SI{20}{\nano\metre} with a cutting current of \SI{1.5}{\nano\A}. 
This resulted in an anisotropic voxel size of $\SI{10}{\nano\metre}\times \SI{10}{\nano\metre}\times\SI{20}{\nano\metre}$. During the tomography, an alternation of the FIB serial sectioning process and the SEM image scanning, 2300 slices were acquired at a resolution of  $4096 \times 3072$  pixels.

\subsection{Image segmentation}\label{sec:imageseg}

In this section, image pre-processing is described. First, in Section~\ref{sec:imageseg_HZB} we explain the procedure by means of which each voxel in the raw FIB-SEM data, acquired as described in Section~\ref{sec:imaging}, is either classified as pore space, additives or active material. This will be called a phase-based segmentation. Subequently, we consider a cutout of the image classified in this way, which contains only additives and pores. In Section~\ref{sec:imageseg_ilastik} the additive phase of this cutout is further segmented, \emph{i.e.}, the graphite particles are separated from the mixture of carbon black and PVDF binder. This further  segmented image will be the basis for data-driven stochastic 3D nanostructure modeling in Section~\ref{sec:modeling}. Possibilities to validate the entire segmentation procedure are briefly discussed in Section~\ref{sec.val.ima}.

\subsubsection{Segmentation of gray scale image: classification of pores, additives and active material}\label{sec:imageseg_HZB}
In the following, we describe the phase-based segmentation of gray scale image data considered in this paper, where a voxel is assigned to the pore phase if, based on its gray value, it is attributed to the silicon-based embedding material, see Section~\ref{sec:imaging}. Due to the tilted nature of the scanned serial sections, the lower part of the sections are located deeper in the sample than the upper regions. 
This change in focal depth was  compensated by adjusting the dynamic focus of the SEM, but the intensity gradient resulting in a darker lower part of the images compared to the upper regions needs still to be compensated. 
For this purpose, each slice of the tomographic image data is normalized by a linear gradient ramp, \emph{i.e.}, a 2D image that has constant values in $x$-direction and linearly changing values in $y$-direction.
This gradient ramp is estimated by first computing the minimum intensity projection of the tomographic image data along the direction of the serial sections. The resulting 2D image was then further smoothed by a Gaussian filter with a standard deviation of 100 pixels (1$\upmu\mathrm{m}$) and a linear ramp was fitted to the smoothed image. 
Finally, the tomographic image data is normalized by dividing the gray scale values of each slice pixelwise by the gray scale values of the ramp.
The entire image stack was then registered and drift corrected using the drift correction based on a scale-invariant feature transform (SIFT)~\cite{lowe.2004}, which is implemented in Fiji~\cite{fiji}. 
As the active material exhibits significantly brighter intensities, it is classified by an automatic Otsu thresholding~\cite{otsu.1979}.

In order to segment the remainder of the image data into pores and additives, a more involved approach is required. Due to the low conductivity of the infiltrating silicon-based resin, local charging artifacts occurred, particularly in areas with low amounts of carbon matrix material. 
Thus, global thresholding does not provide an appropriate binarization. 
Therefore, we perform a slicewise segmentation using the 2D Weka segmentation~\cite{arganda.2017} as in~\cite{osenberg.2022}, which is trained with hand-labeled data.
In the segmented data, physically unrealistic effects are observed in terms of isolated voxels of the additive phase or small pore inclusions. 
These effects are removed by applying a morphological closing  to the additive phase. As structuring element for the morphological closing~\cite{soille.2003}, a rectangular cuboid of a size of one voxel is used. Finally, all parts of the additive phase, which still belong to isolated components after morphological closing, are removed. For this purpose, we use the connected component analysis implemented in the Fiji software package morpholibJ~\cite{legland.2016}.
For subsequent image processing and model fitting, we performed a down-sampling on the tomographic image data such that we result in cubic voxels with equal side lengths of $\SI{20}{\nano\metre}.$

\subsubsection{Segmentation of additives: classification of graphite particles}\label{sec:imageseg_ilastik}
\begin{figure}[h]
    \centering
    \includegraphics[width=\textwidth]{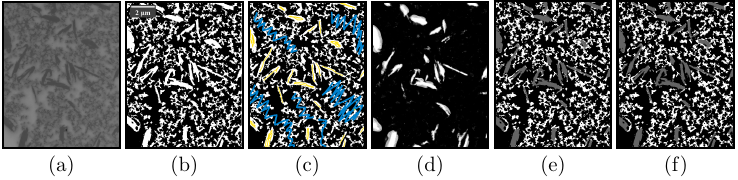}
    \caption{Visualization of image segmentation procedure at the example of a 2D slice: Raw FIB-SEM image data~(a). Binarized  image data, where the binder additive phase and the pore space are represented in white and black, respectively~(b). Hand-labeled image data, which serves as input for training a random forest. Graphite particles are labeled in yellow, while its complement is labeled in blue~(c). Probability map predicted by the random forest trained with the software ilastik. For voxels with brighter gray scale values, a larger probability is predicted for belonging to a graphite particle~(d). Segmented  image data obtained from the probability map~(e). Segmented image data after final post-processing~(f).}
    \label{fig:ilastik}
\end{figure}

From now on, we consider a 3D cutout of the tomographic image data that does not contain any active material. The cutout is a rectangular cuboid of size $\SI{9.76}{\micro\metre}\times \SI{14.92}{\micro\metre}\times \SI{28.6}{\micro\metre}.$ 
The additive phase in this sample consists of graphite particles, carbon black and PVDF binder. Based on their size and shape, the graphite particles can be distinguished from the union of carbon and PVDF binder. 
While the latter forms a fine-grained homogeneous structure, compared to which the oblate-shaped and dense graphite particles are coarse, see Figure~\ref{fig:ilastik}b.
Due to the different morphology of graphite particles on the one hand and the mixture of carbon black and PVDF binder on the other hand, we account for them separately in the stochastic 3D  model introduced in Section~\ref{sec:modeling}. Thus, as a basis for model calibration, we perform a segmentation of image data which further subdivides the additive phase into graphite particles and the mixture of carbon black and PVDF binder.

For this purpose, we proceed as follows. Note that after the segmentation performed in Section~\ref{sec:imageseg_HZB}, it is sufficient to separate the graphite particles from the rest of the image. First, we hand-label the graphite particles as well as their complement on a few 2D slices of the tomographic image data, see Figure~\ref{fig:ilastik}c. 
The segmentation is then carried out by a random forest~\cite{breiman2001random} that classifies each voxel based on features which contain information about the morphology in the neighborhood of the given voxel. As features, we consider a Gaussian smoothing ($\sigma = 5.0$), the magnitude of the gradient after Gaussian smoothing ($\sigma = 15.0$), the eigenvalues of the structure tensor ($\sigma=3.5$, $\sigma = 5.0$) and the eigenvalues of the Hessian matrix after Gaussian smoothing ($\sigma=1.6$, $\sigma = 5.0$), where the values of $\sigma$ are given in voxel units. 
The output is a probability map that assigns each voxel a probability of belonging to a graphite particle, see Figure~\ref{fig:ilastik}d. 
The probability map is then binarized by thresholding it at the $85\%$-quantile of its distribution function, which results in a binary 3D image that labels the locations of voxels belonging to a graphite particle. We choose a quantile of $85\%$ rather than $50\%$ as we are only interested in regions that are detected as graphite particles with high confidence.
We use this binarized probability map to label additive voxels as either graphite or the mixture of carbon black and PVDF binder, while keeping the pore voxels unchanged, see Figure~\ref{fig:ilastik}e.

We can observe that there is a small amount of granular misclassification in the fine-grained phase.
Furthermore, after the segmentation, some large graphite particles are not properly identified along their edges or exhibit small and unrealistic holes inside. In order to improve the segmentation with respect to these aspects, two post-processing steps are performed. First, we compute the connected components of the graphite phase, see Figure~\ref{fig:ilastik}e, and remove those connected components that have a volume of less than 5000 voxels ($\approx$ \SI{0.04}{\micro\metre\cubed}), which corresponds to the volume of a ball with diameter of 10.6 voxels ($=$ \SI{212}{\nano\metre}). Doing so, unrealistically small regions are removed from the graphite phase that are probably misclassified.  
Second, we perform a morphological closing~\cite{soille.2003} on the segmented graphite phase of the labeled image where the structuring element is a ball with radius $10$ voxels ($=$ \SI{200}{\nano\metre}), which closes small gaps within the graphite phase and smooths its edges.

With this procedure it might happen that voxels that are classified as pore space in Section~\ref{sec:imageseg_HZB}, are now classified as graphite particles. We correct for this as follows. Voxels that are classified as additives in Section~\ref{sec:imageseg_HZB} and as graphite in the present section are finally classified as graphite, while voxels that were classified as pore space in Section~\ref{sec:imageseg_HZB} remain classified as pore space, even if the random forest classifier described in this section labels them as graphite. The final result is visualized in Figure~\ref{fig:ilastik}f. The entire segmentation process, including hand labeling, feature selection and training of the random forest classifier, was done with the software ilastik~\cite{berg.2019}. All other image processing steps were done with Fiji~\cite{fiji}.

\subsubsection{Validation of image segmentation}\label{sec.val.ima}

It is possible to validate the  image segmentation procedure  based on the composition of the cathode described in Section~\ref{sec:sample_preparation}. Using estimated densities of each constituent (\SI{4.77}{\gram\per\centi\metre\cubed} for active material~\cite{mueller.2021}, \SI{1.8}{\gram\per\centi\metre\cubed} for PVDF binder (manufacturer information), \SI{2.3}{\gram\per\centi\metre\cubed} for graphite particles (pycnometer-measurement), \SI{1.9}{\gram\per\centi\metre\cubed} for carbon black (pycnometer-measurement)) and the porosity of a subdomain with no active material, which is estimated to $66.86\%$ from the image segmentation obtained in Section~\ref{sec:imageseg_HZB}, we can compute the volume fractions of graphite, carbon black, and PVDF binder within a subdomain of the material that does not contain any active material. 
This yields a theoretical volume fraction of $11.08\%$ for graphite, $10.73\%$ for carbon black and $11.33\%$ for PVDF binder. 
On the other hand, the image segmentation procedure described in this section attributes $10.55\%$ of the voxels to  graphite particles and  $22.59\%$ of the voxels to the mixture of  PVDF binder and carbon black. These values are close to those obtained based on the manufacturing process and the densities of the constituents, which validates our segmentation visualized in Figure~\ref{fig:ilastik}f. Note that this can also been considered as an \emph{a posteriori} validation of the segmentation presented in Section~\ref{sec:imageseg_HZB}, which serves as a basis for the segmentation considered in Section~\ref{sec:imageseg_ilastik}.

\section{Stochastic nanostructure modeling}\label{sec:modeling}

In Section~\ref{sec:model_desc},
we give a detailed description of the stochastic nanostructure model introduced in this paper.  Then, in Section~\ref{sec:modelcalib}, we describe how the model parameters  are estimated from tomographic image data, whereas model validation is performed in Section~\ref{sec:modelval} by comparing various morphological descriptors that are not used for model fitting, computing them for model realizations and tomographic image data, respectively. The framework of a simulation study for varying amounts of graphite is explained in Section~\ref{sec:vary_graphite}.

\subsection{Model description}\label{sec:model_desc}
The main idea of the stochastic 3D  model for the additive phase is to combine excursion sets of Gaussian random fields with Boolean models. First, in Section~\ref{sec.boo.gra}, the graphite particles are modeled with a Boolean model, where the grains are oblate spheroids with random size and orientation. In the second step, the mixture of PVDF binder and carbon black is modeled in the complement of the Boolean model by an excursion set of a Gaussian random field, see Section~\ref{sec.exc.set}. Subsequently, in Section~\ref{sec.boo.por}, a second Boolean model with spherical grains is used to thin the initial excursion set by removing all virtual PVDF binder and carbon black from areas that are covered by this  Boolean model. Note that this modeling step allows for representing the inhomogeneous distribution of PVDF binder and carbon black more accurately. 
Finally, in Section~\ref{sec.com.sto},  it is shown how the three models can be combined to get a stochastic 3D model for the entire additive phase.

\subsubsection{Boolean model for graphite particles}\label{sec.boo.gra}
In this section, we briefly describe the Boolean model which is used to model the union of graphite particles (gray phase  in Figure~\ref{fig:ilastik}f). For a comprehensive introduction to this kind of spatial stochastic models, we refer to   Chapters~2 and 3 of~\cite{chiu.2013}. 
Let $X = \{X_1, X_2, \ldots\}$ denote a homogeneous Poisson point process in $\R^3$ with some intensity $\lambda_X>0$, where  $X_1, X_2, \ldots$ are random vectors with values in $\R^3$, which are called the \emph{germs} of the model.
Moreover, let $E_1,E_2,\ldots$ be a sequence of independent and identically distributed random closed sets in $\R^3$,  so-called  \emph{grains}. 
Due to the morphology of graphite particles observed in 3D image data, we assume that the grains are isotropic random oblate spheroids centered at the origin.
Note that an oblate spheroid that is centered at the origin can be defined by the lengths $a,c>0$ of  two half-axes, where the length of the third half-axis is  equal to $\max \{a,c \}$. The random  lengths $A_1,A_2,\ldots$ and $C_1,C_2,\ldots$ of the half-axes of the random spheroids $E_1,E_2,\ldots$ are chosen to be independent random variables, which are gamma distributed  with shape parameters $\alpha_1,\alpha_2>0$ and rate parameters $\gamma_1,\gamma_2>0$, respectively. To keep the model simple, we assume that the rate parameters are identical, i.e. $\gamma_1=\gamma_2 = \gamma$ for some $\gamma > 0$. This reduces the total number of parameters, which--in turn--increases stability and reproducibility of the numerical algorithm used for estimating the model parameters as described in Section~\ref{sec:modelcalib} below. On the other hand,  the model remains sufficiently flexible in order to describe the  union of graphite particles  with the required accuracy.

The Boolean model, denoted by  $\Xi^{(1)}$, is then defined as the union of the random oblate spheroids $E_1,E_2,\ldots$  shifted by the random germs $X_1,X_2,\ldots$, \emph{i.e.}
\begin{equation}
    \Xi^{(1)} =\bigcup_{i=1}^{\infty} \left(E_i+X_i\right),
\end{equation}
where $E_i+X_i=\{y+X_i:  y\in E_i\}$ for each $i\in\N=\{1,2,\ldots\}$.  
Note that the shifted grains $E_i+X_i$ and $E_j+X_j$  can overlap for some $i,j\in\N$ with $j\neq i$. 
This is a desired effect, as--by the union of two or more spheroidal grains--it allows the generation of graphite particles whose shapes differ from oblate spheroids.

\subsubsection{Excursion set model for the mixture of carbon black and binder}\label{sec.exc.set}

In order to model the mixture of binder and carbon black (white phase in Figure~\ref{fig:ilastik}f) we use the following approach. Let $Z=\{Z(t)\colon t\in \R^3\}$ be a motion invariant, \emph{i.e.}, stationary and isotropic, Gaussian random field, which is independent of the Boolean model $\Xi^{(1)}$ introduced in Section~\ref{sec.boo.gra}. Moreover, assume that $\operatorname{\mathbb{E}}[Z(t)]=0$ and $\operatorname{Var}[Z(t)]=1$ for each $t\in\R^3$. The covariance function of $Z$ will be denoted by $\rho\colon \R^3\times \R^3 \rightarrow \R$. For an introduction to Gaussian random fields and their geometric properties, we refer to \cite{chiu.2013, lantuejoul.2002, adler.1981}. Note that by the motion invariance of $Z$, the value of $\rho(s,t)$ depends only on the length $\vert s-t\vert$ of the vector $s-t$ for any $s,t\in \R^3$.
Hence, we write $\rho(h)=\rho(s,t)$ for any $h\geq 0$ and $s,t\in\R^3$  with $h=\vert s-t \vert$.
Under certain regularity conditions on the random field $Z$, the excursion set $\Xi^{(2)}=\{t\in\R^3\colon Z(t)\geq \mu\}$ is a random closed set in $\R^3$ for each $\mu \in \R$, see Section 5.2.1 in~\cite{molchanov.2005}, where $\Xi^{(2)}$ is also called the $\mu$-excursion set of the random field $Z$. 
As we will see  later on in Section~\ref{sec:gauss_calib}, to describe the mixture of binder and carbon black by $\Xi^{(2)}$, it turns out that the covariance function $\rho$ can be approximated well through a parametric fit by assuming that $\rho$ is of the form 
\begin{equation}\label{eq:rho_parametric_form}
    \rho(h) = \frac{1}{1+(\eta h)^2},
\end{equation}
for each $h>0$, where $\eta >0$ is some parameter.

\subsubsection{Boolean model for large pores}\label{sec.boo.por}

Finally, to describe the union of large pores, we still consider another Boolean model, denoted by $\Xi^{(3)}$, which is given as follows. Let $Y=\{Y_1,Y_2,\ldots\}$ be a homogeneous Poisson point process in $\R^3$ with intensity $\lambda_Y>0$ and let $R_1,R_2,\ldots$ be a sequence of independent and (identically) exponentially  distributed random variables  with rate parameter $\theta>0$.
Furthermore, let $\Xi^{(3)}$ be a random set in $\R^3$ which is the union of spherical grains whose radii are given by $R_1, R_2, \ldots$, \emph{i.e.},
\begin{equation}
    \Xi^{(3)}=\bigcup_{i=1}^\infty B(Y_i, R_i),\label{lab.equ.thr}
\end{equation} 
where $B(Y_i, R_i)=\{x\in\R^3: \vert x-Y_i\vert<R_i\}$ denotes the \emph{open} ball with radius $R_i$ centered at $Y_i$ for $i\in\N$. Note that strictly speaking, the random set $\Xi^{(3)}$ considered in Eq.~\eqref{lab.equ.thr} does not constitute a Boolean model as it is not a random closed set. However, the set $\Xi^{(3)}$ will enter our final stochastic 3D model, stated in Section~\ref{sec.com.sto} for the entire additive phase, only through its complement $\big(\Xi^{(3)}\big)^{\text{c}}$. We therefore choose open balls $B(Y_i, R_i)$ 
in Eq.~\eqref{lab.equ.thr}
to ensure that $\big(\Xi^{(3)}\big)^{\text{c}}$ is closed.

\subsubsection{Combined stochastic 3D model for the entire additive phase}\label{sec.com.sto}

Instead of considering the random excursion set $\Xi^{(2)}$ introduced in Section~\ref{sec.exc.set}, we  intersect it with $\big(\Xi^{(3)}\big)^c$. 
This means that two further parameters (namely, $\lambda_Y$ and $\theta$)  have to be taken into account, in addition to the model parameters of $\Xi^{(2)}$. However, this modification of $\Xi^{(2)}$ allows for a better representation of the heterogeneous spatial distribution of carbon black and PVDF binder by creating additional pore regions without altering the shape of the graphite particles. The final stochastic 3D model, denoted by $\Xi$, for the entire additive phase  (including graphite particles, carbon black, PVDF binder) is then given by \begin{equation}
    \Xi = \Xi^{(1)} \cup \big(\Xi^{(2)}\cap \big(\Xi^{(3)}\big)^c\big).\label{lab.equ.fou}
\end{equation}
Note that since the random closed sets $\Xi^{(1)}$, $\Xi^{(2)}$ and $\big(\Xi^{(3)}\big)^c$ are  motion invariant, the random closed set $\Xi$ considered in Eq.~\eqref{lab.equ.fou} is also motion invariant.

\subsubsection{Physical interpretation of model parameters}
All parameters of our stochastic nanostructure model $\Xi$, i.e. $\lambda_X$, $\alpha_1$, $\alpha_2$, $\gamma$, $\mu$, $\eta$, $\theta$ and $\lambda_Y$, allow for a physical interpretation. In the Boolean model $\Xi^{(1)}$ mimicking the graphite particles, the parameter $\lambda_X$ is the expected number of individual ellipsoidal grains per unit volume, while $\alpha_1$, $\alpha_2$ and $\gamma$ determine size and elongation of the graphite particles. The latter are modeled as oblate spheroids, where the desired mean and variance of their half-axis length are controlled by $\alpha_1$, $\alpha_2$ and $\gamma$. For the excursion set $\Xi^{(2)}$ modeling the mixture of carbon binder and additives, the volume fraction can be adjusted by the parameter $\mu$, while the parameter $\eta$ of the covariance function represents a scale parameter controlling how fine-grained the structure of the resulting set $\Xi^{(2)}$ is. Larger values of $\eta$ lead to finer structures. Finally, for the Boolean model $\Xi^{(3)}$ which represents large pore regions, the parameter $\lambda_Y$ is the expected number of spherical grains per unit volume, while $\theta$ is the inverse of the mean radius of the spherical grains. With these two parameters, we can vary the morphology of large pore regions.

\subsection{Model calibration}\label{sec:modelcalib} 
In this section we explain how the model parameters of 
the random closed set $\Xi$ considered in Eq.~\eqref{lab.equ.fou} are calibrated, i.e.,
how the parameter vector 
\begin{equation}
(\lambda_X,\alpha_1,\alpha_2,\gamma, \eta,\theta,\lambda_Y, \mu)\in (0, \infty)^7 \times \R
\end{equation}
is estimated, based on the segmented tomographic image data presented in Section~\ref{sec:materials}.

\subsubsection{Boolean model for graphite particles}\label{sec:est:BooleanModel1}
We first explain how the Boolean model $\Xi^{(1)}$, describing the union of  oblate shaped (graphite) particles, is calibrated. The basis of this calibration is the gray phase in the segmented tomographic image data  presented in Section~\ref{sec:materials}, see Figure~\ref{fig:ilastik}f.
Here, the intensity $\lambda_X$ of the underlying Poisson point process as well as the parameters $\alpha_1$, $\alpha_2$, $\gamma>0$  of the gamma distributions modeling the lengths of the half-axes  of the oblate spheroids have to be estimated from image data. 

The values of these four parameters are optimized in order to minimize the discrepancy between model and data with respect to the densities of four different intrinsic volumes~\cite{chiu.2013, ohser.2009} of the stationary random closed set $\Xi^{(1)}$, \emph{i.e.}, the volume fraction, the surface area per unit volume, the specific integral of mean curvature and the specific connectivity number, which are denoted by $V_1$, $S_1$, $K_1$ and $N_1$, respectively. Provided that the boundary of $\Xi^{(1)}$ is almost surely sufficiently smooth almost everywhere, they are defined by 
\begin{align}
V_1 &= \frac{1}{ \nu_3(W)}\,\mathbb{E}[\nu_3(\Xi^{(1)} \cap W)] \\
S_1 &= \frac{1}{ \nu_3(W)}\,\mathbb{E}[\mathcal{H}_{2}(\partial \Xi^{(1)} \cap W)], \\
K_1 &= \frac{1}{ \nu_3(W)}\,\mathbb{E}\left[ \int_{\partial \Xi^{(1)} \cap W} H_{1}(x) \mathcal{H}_{2} (\mathrm{d}x) \right],\\
N_1 &= \frac{1}{\mathcal{H}_{2}\big(\partial B(o,1)\big) \nu_3(W)}\,\mathbb{E}\left[ \int_{\partial \Xi^{(1)} \cap W} H_{2}(x) \mathcal{H}_{2} (\mathrm{d}x) \right],
\end{align} 
for some Borel-measurable $W \subset \R^3$ with $\nu_3(W) > 0$, where $o, \nu_3$, $\mathcal{H}_{2}, H_{1}(x), H_{2}(x)$ and $\partial \Xi$ denote the origin in $\R^3$, the three-dimensional Lebesgue measure, the two-dimensional Hausdorff measure, the mean curvature of $\Xi^{(1)}$ in $x$, the Gaussian curvature of $\Xi^{(1)}$ in $x$ and the boundary of $\Xi^{(1)},$ respectively. Definitions for the mean and Gaussian curvatures can be found in~\cite{ohser.2009}. Note that the definition of each intrinsic volume does not depend on the actual choice of $W$. These quantities can be estimated from the tomographic image data using the algorithms described in~\cite{ohser.2009}. 

For a given set of a parameters $\lambda_X,\alpha_1,\alpha_2,\gamma$, we can  compute the values of $V_1$, $S_1$, $K_1$ and $N_1$ using Miles' formulas, see Eqs.~(3.45)--(3.48) in~\cite{chiu.2013}. Namely, it holds that
\begin{align}\label{eq:miles1}
    V_1 &= 1-\exp(-\lambda_X \overline{V}), \\
    S_1 &= \lambda_X (1-V_1) \overline{S},\\
    K_1 &= \lambda_X (1-V_1)\biggl(\overline{K}-\frac{\pi^2\lambda_X\overline{S}^2}{32}\biggr),\\
    N_1 &= \lambda_X (1-V_1)\biggl(1-\frac{\lambda_X\overline{K}\bar{S}}{4\pi}+\frac{\pi\lambda_X^2\overline{S}^3}{384}\biggr),\label{eq:miles4}
\end{align}
where $\overline{V}$, $\overline{S}$ and $\overline{K}$ denote the expected volume, expected surface area and expected integral of mean curvature, respectively, of the individual random grains $E_1,E_2,\ldots$. Furthermore, since in our case the grains are oblate spheroids with random half-axes lengths $A_1,A_2,\ldots$ and $C_1,C_2,\ldots$, the following  formulas for  $\overline{V}$, $\overline{S}$ and $\overline{K}$ can be used:
\begin{align} \label{eq:spheroidvol1}
    \overline{V} &= \mathbb{E}\left[\frac{4}{3}\pi \delta \max\{A_1^3,C_1^3\}\right], \\
    \overline{S} &= \mathbb{E}\left[2\pi \delta \max\{A_1^2,C_1^2\} \left(\frac{1}{\delta}-\frac{\delta}{\sqrt{1-\delta^2}}\ln \frac{1-\sqrt{1-\delta^2}}{\delta}\right)\right],\\\label{eq:spheroidvol3}
    \overline{K} &= \mathbb{E}\left[2\pi \max\{A_1,C_1\} \left(\delta + \frac{1}{\sqrt{1-\delta^2}}\arcsin \sqrt{1-\delta^2}\right)\right],
\end{align} 
where $\delta=\min\{\frac{A_1}{C_1},\frac{C_1}{A_1}\} \leq 1$,
see Table~1.1 and Eq.~(1.27) in~\cite{chiu.2013}.

While the right-hand sides of Eqs.~\eqref{eq:spheroidvol1}--\eqref{eq:spheroidvol3} can easily be computed from a given realization of $\Xi^{(1)}$, these equations only give an implicit dependence of $\overline{V}$, $\overline{S}$ and $\overline{K}$ on the model parameters $\lambda_X$, $\alpha_1$, $\alpha_2$ and $\gamma$, which control the distribution of $A_1$ and $C_1$.
%
%
In order to obtain an explicit dependence on $\lambda_X$, $\alpha_1$, $\alpha_2$ and $\gamma$, we employ a numerical approach, where the expectations in Eqs.~\eqref{eq:spheroidvol1}--\eqref{eq:spheroidvol3} are approximated through the strong law of large numbers. 
For this, we independently draw 10000 realizations from the distribution of $A_1$ and $C_1$, respectively,  and insert them into the expressions within the brackets in Eqs.~\eqref{eq:spheroidvol1}--\eqref{eq:spheroidvol3}. By averaging over the resulting values, we get
 consistent estimators for $\overline{V}$, $\overline{S}$ and $\overline{K}$. Using Eqs.~\eqref{eq:miles1}--\eqref{eq:miles4} we can then compute the values of the densities $V_1$, $S_1$, $K_1$ and $N_1$ of the Boolean model $\Xi^{(1)}$ with given parameters $\lambda_X,\alpha_1,\alpha_2,\gamma$.
 
This allows us to numerically determine a set of parameters  $\lambda_X,\alpha_1,\alpha_2,\gamma$ such that the corresponding values of the densities $V_1$, $S_1$, $K_1$ and $N_1$
closely match the ones estimated from segmented image data. For this we used the Nelder-Mead method~\cite{nelder.1965}, which is a numerical gradient-free search method, to solve the minimization problem
\begin{equation}\label{eq:minimization}
    \min_{(\lambda_X,\alpha_1,\alpha_2,\gamma)}(V_1 - \widehat{V_1})^2+(20~\unit{\nano\metre})^2(S_1-\widehat{S_1})^2+(20^2~\unit{\nano\metre\squared})^2(K_1-\widehat{K_1})^2+(20^3~\unit{\nano\metre\cubed})^2(N_1-\widehat{N_1})^2,
\end{equation}
where $\widehat{V_1}$, $\widehat{S_1}$, $\widehat{K_1}$ and $\widehat{N_1}$ denote the estimated densities of intrinsic volumes, which are computed from tomographic image data by means of the algorithms described in~\cite{ohser.2009}. We used the implementation of the
Nelder-Mead method given in the Matlab function \texttt{fminsearch}, using the methodology described in~\cite{fminsearch}. 
The values obtained from this optimization procedure are shown in Table~\ref{tab:intrisic_vols} alongside the target values of $\widehat{V_1}$, $\widehat{S_1}$, $\widehat{K_1}$ and $\widehat{N_1}$ obtained from tomographic image data.

\begin{table}[h]
    \centering
\begin{tabular}{lcccc}
              & $\widehat{V_1}$   & $\widehat{S_1}$~[\unit{\per\nano\metre}]    & $\widehat{K_1}$~[\unit{\per\nano\metre\squared}]                 & $\widehat{N_1}$~[\unit{\per\nano\metre\cubed}]        \Tstrut\Bstrut\\\hline
tomographic image data    & 0.10550 & 0.0014526  & $4.2043\cdot 10^{-7}$ & $\phantom{-}1.6344\cdot 10^{-9}$\Tstrut\Bstrut \\\hline
 & $V_1$   & $S_1$~[\unit{\per\nano\metre}]    & $K_1$~[\unit{\per\nano\metre\squared}]                 & $N_1$~[\unit{\per\nano\metre\cubed}]        \Tstrut\Bstrut\\\hline
calibrated Boolean model &  0.10569 & 0.0014377 & $4.5445\cdot 10^{-7}$ & $-1.3038\cdot 10^{-9}$\Tstrut\Bstrut            
\end{tabular}
    \caption{Comparison of densities 
    $\widehat{V_1}$, $\widehat{S_1}$, $\widehat{K_1}, \widehat{N_1}$ and $V_1$, $S_1$, $K_1,N_1$
    of intrinsic volumes of graphite particles, computed for tomographic image data and the calibrated Boolean model  $\Xi^{(1)}$, respectively.}
    \label{tab:intrisic_vols}
\end{table}

\subsubsection{Excursion set model for the mixture of carbon black and binder}\label{sec:gauss_calib}
To calibrate the excursion set model $\Xi^{(2)}$ introduced in Section~\ref{sec.exc.set}, 
we need to estimate the covariance function $\rho$ of the underlying Gaussian random field $Z$, as well as the threshold $\mu\in\R$ at which the random field $Z$ is truncated.
For this, we first choose a cutout of the tomographic image data that shows a representative distribution of carbon black and binder without graphite particles and large pores. The size of this cutout is $278\times 136\times 124$ voxels ($\SI{5.56}{\micro\metre}\times\SI{2.72}{\micro\metre}\times\SI{2.48}{\micro\metre}$), see Figure~\ref{fig:app_gaussian_cutout} of the appendix. In order to estimate $\mu$, note that \begin{equation}\label{eq.vau.phi}
    V_2=1-\Phi(\mu),
\end{equation}
where $V_2=P(o\in\Xi^{(2)})$ is the volume fraction of the stationary random closed set $\Xi^{(2)}$, and $\Phi\colon\R\to[0,1]$ is the cumulative probability distribution function of the standard normal distribution. Thus, we first determine 
an estimator $\widehat{V_2}$ for the volume fraction of $V_2$ based on the above-mentioned cutout of the tomographic image data, proceeding
analogously as in Section~\ref{sec:est:BooleanModel1}. Then, using Eq.~\eqref{eq.vau.phi}, a so-called plug-in estimator $\widehat{\mu}$ for $\mu$ is  given by
\begin{equation}
    \widehat{\mu}=\Phi^{-1}(1-\widehat{V_2}).
\end{equation}

\begin{figure}[h]
    \centering
    \includegraphics[width=0.4\textwidth]{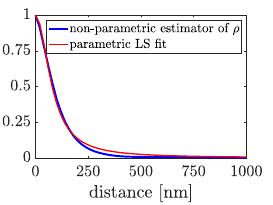}
    \caption{Comparison of non-parametric and parametric estimates for the covariance function $\rho$ of the Gaussian random field $Z$.}
    \label{fig:cov_estimates}
\end{figure}

In order to estimate the covariance function $\rho$, we consider the two-point coverage probability function $C\colon\R^3\times\R^3\rightarrow\R$ of $\Xi^{(2)}$ defined by
\begin{equation}\label{eq:twopointcovdef}
\begin{aligned}
	C(s,t) =  P(s\in\Xi^{(2)},t\in\Xi^{(2)})
\end{aligned}
\end{equation}
for any $s,t\in\R^3$.
Note that, as for the covariance function $\rho$, the motion invariance of $Z$ allows us to consider $C$ as a mapping from $[0,\infty)$ to $\R$ where, for each $h \geq 0$, we have $C(h)=C(s,t)$ for arbitrary $s,t\in\R^3$ such that $\vert s-t\vert  = h$. We then make use of the fact that
\begin{equation}\label{eq:twopointcoverage}
    C(h)=V_2^2+\int_0^{\rho(h)}\displaystyle\frac{e^{\frac{-\mu^2}{1+
    t}}}{\sqrt{1-t^2}}\,\textup{d}t,
\end{equation}
for each $h>0$, see Proposition 16.1.1 of~\cite{lantuejoul.2002}, and that the two-point coverage probability $C(h)$ can easily be  estimated from image data as described in Section 6.2.3 of~\cite{ohser.2009}.
Note that for any $h>0$, the right-hand side of~\eqref{eq:twopointcoverage} is strictly increasing in $\rho(h)$.
Thus, after replacing $C(h)$, $V_2$ and $\mu$ with their respective estimators, we can solve Eq.~\eqref{eq:twopointcoverage} for $\rho(h)$ numerically, using the method of bisection~\cite{Burden1989}. This yields a preliminary (non-parametric) estimator for $\rho$. In the next step,
as mentioned in Section~\ref{sec:model_desc}, we  determine a parametric model fit to this estimator by assuming that $\rho$ is of the form given in Eq.~\eqref{eq:rho_parametric_form}, where the parameter $\eta >0$ is estimated by a least-squares (LS) approach. It turned out that both estimates  for the covariance function $\rho$ of the Gaussian random field $Z$ are quite similar to each other, see Figure~\ref{fig:cov_estimates}.

\subsubsection{Boolean model for large pores}\label{sec:poissonthinning} In order to calibrate the parameters of the random set $\Xi^{(3)}$ introduced in Section~\ref{sec.boo.por}, we need to estimate the intensity $\lambda_Y>0$ of the underlying Poisson point process and the rate parameter $\theta>0$ of the exponential distribution modeling the random radii of (spherical) grains.
Since the stationary random sets $\Xi^{(1)}$, $\Xi^{(2)}$ and $\Xi^{(3)}$ are assumed to be independent,  the volume fraction $V$ of the combined stochastic 3D model $\Xi$ given in Eq.~\eqref{lab.equ.fou}
can be expressed by the volume fractions $V_1, V_2, V_3$ of $\Xi^{(1)}, \Xi^{(2)}$ and $\Xi^{(3)}$.  Namely, it holds that
\begin{equation}\label{eq:poissonthin1}
V=V_1 + V_2(1-V_1)(1-V_3),
\end{equation}
because
\begin{equation}
    \begin{aligned}
    V &= P(o\in\Xi)=P\left(o\in \Xi^{(1)}\cup\left(\Xi^{(2)}\cap (\Xi^{(3)})^c\right)\right)\\
    &=P\left(o\in \Xi^{(1)}\right) + P\left(o\in\Xi^{(2)}\cap (\Xi^{(3)})^c\right) - P\left(o\in\Xi^{(1)}\cap \Xi^{(2)}\cap (\Xi^{(3)})^c\right)\\
    &=P\left(o\in \Xi^{(1)}\right) + P\left(o\in \Xi^{(2)}\right)\left(1-P(o\in \Xi^{(1)})\right)\left(1 - P\left(o\in \Xi^{(3)}\right) \right)\\
    &=V_1 + V_2(1-V_1)(1-V_3).
\end{aligned}
\end{equation}
The volume fractions $V_1$ and $V_2$ of $\Xi^{(1)}$ and $\Xi^{(2)}$ have already been determined, as explained in the previous sections, and the volume fraction $V$ of $\Xi$ can easily be estimated from tomographic image data.
Moreover, using  Miles' formula for the volume fraction of stationary Boolean models given in Eq.~\eqref{eq:miles1},  we get that 
\begin{equation}\label{eq:poissonthin5}
    V_3=1-\exp\left(- \lambda_Y \, \frac{8\pi}{\theta^3}\right),
\end{equation}
 where we used that $\mathbb{E} [R_1^3] = 6 / \theta^3$. Combining Eq.~\eqref{eq:poissonthin1} with Eq.~\eqref{eq:poissonthin5} leads to 
 \begin{equation}\label{eq:poissonthin_intensity}
 	\lambda_Y = -\frac{\theta^3}{8\pi} \log \left( \frac{V- V_1}{V_2 (1 - V_1)}\right).
 \end{equation}
Note that the rate parameter $\theta>0$ of the exponential radius distribution of the spherical grains of $\Xi^{(3)}$ is the only remaining unknown parameter on the right-hand side of Eq.~\eqref{eq:poissonthin_intensity}.
Thus, for a given value of $\theta$, the corresponding value of $\lambda_Y$ is determined by means of Eq.~\eqref{eq:poissonthin_intensity}, ensuring that the volume fraction of the final model $\Xi$ matches the one observed in tomographic image data.

Our goal is to choose the value of $\theta$ in such a way that the difference between the pore space morphology of realizations of the combined stochastic 3D model $\Xi$ and tomographic image data is minimized. 
For this, we consider the continuous pore size distribution  $\operatorname{CPSD}\colon [0,\infty)\rightarrow [0,1]$ of $\Xi$, which is defined as follows.
For each $r\geq 0$, the value of $\operatorname{CPSD}(r)$ is equal to the  fraction of the pore space of $\Xi$ that can be covered by spheres of radius $r$ which  are completely contained within the pore space. Formally, it holds that
\begin{equation}\label{eq:CPSD_def}
\operatorname{CPSD}(r) =  \frac{\mathbb{E}\bigg[\nu_{3}\bigg( \Big(\big(\Xi^c\ominus B(o,r)\big)\oplus B(o,r)\Big) \cap [0,1]^3   \bigg)\bigg]}{\mathbb{E}\left[ \nu_{3}\big(\Xi^c\cap [0,1]^3\big)\right]},
\end{equation}
for each $r\geq 0$, where $\ominus$ and $\oplus$ denote morphological erosion and dilation~\cite{soille.2003,serra.1982}, respectively, $B(o,r)=\{x\in\R^3\colon \vert x\vert <r\}$ is the open ball with radius $r$ centered at the origin $o\in\R^3$, and $\nu_3$ denotes the three-dimensional Lebesgue measure.
Note that the value of $\operatorname{CPSD}(r)$ for $r\geq 0$ can easily be computed from voxelized image data using the Euclidean distance transform~\cite{soille.2003,maurer.2003}.

By generating model realizations for different values of $\theta$,
we can now heuristically minimize the $L_1$-distance between the continuous pore size distributions  of model realizations and tomographic image data.
To be precise, we generate three model realizations for each $\theta\in\{0.0125, 0.0118, 0.0111, 0.0105, 0.0100, 0.0095, 0.0091\}$~\unit{\per\nano\metre}
and compare the resulting continuous pore size distributions with the continuous pore size distribution computed from tomographic image data, see Figure~\ref{fig:app_nu_choice} of the appendix.
The  values of all parameters of the calibrated  model $\Xi$ are shown in Table~\ref{tab:est_parameters}.
\begin{table}[h]
    \centering
    \begin{tabular}{cccc|cc|cc}
    \multicolumn{4}{c|}{$\Xi^{(1)}$} & \multicolumn{2}{c|}{$\Xi^{(2)}$} &  \multicolumn{2}{c}{$\Xi^{(3)}$} \Tstrut\Bstrut\\\hline
    $\lambda_X$[\unit{\per\nano\metre\cubed}] & $\alpha_1$[\unit{\nano\metre}] & $\alpha_2$[\unit{\nano\metre}] & $\gamma$[-] & $\mu$[-] & $\eta$[\unit{\per\nano\metre}] & $\theta$[\unit{\per\nano\metre}] & $\lambda_Y$[\unit{\per\nano\metre\cubed}]\Tstrut\Bstrut\\\hline
    6.355$\cdot 10^{-11}$ & 205 & 3944 & 1.971 & 0.499 & 0.0127 & 0.0105 & $9.340\cdot 10^{-9}$ \Tstrut\Bstrut
    \end{tabular}
    \caption{Parameter values of the calibrated stochastic 3D model $\Xi$ determined by means of tomographic image data. 
    }
    \label{tab:est_parameters}
\end{table}

\subsection{Model validation}\label{sec:modelval} For a visual comparison of tomographic image data and a realization of the calibrated stochastic  model $\Xi$, see Figure~\ref{fig:3Dvis}, where  one can observe that the model realization nicely reproduces the morphology of the additive phase observed in measured 3D  data.
In the following, we provide a quantitative model validation by comparing various morphological descriptors of the tomographic image data to those of model realizations. For the latter, we draw five  samples from the calibrated stochastic model $\Xi$ which are realized in a cubic sampling window consisting of $800^3$ voxels, where the voxels are cubic with a side length of $\SI{20}{\nano\metre}$.

Since, later on, the generated samples of $\Xi$ serve as input for  numerical simulations of effective physical properties, we also validate the model by considering realizations at voxel resolutions that are coarser than the one used for model calibration. 
For these coarser voxel resolutions, model realizations are generated in a sampling window of the same  size, \emph{i.e.}, $400^3$ and $200^3$ voxels for voxel side lengths of $\SI{40}{\nano\metre}$ and $\SI{80}{\nano\metre}$, respectively.
The motivation behind considering coarser voxel resolutions is that many memory-intensive numerical computations require coarser resolutions in order to be feasible.

We also investigate the influence of down-sampling on morphological descriptors. 
Recall from Section~\ref{sec:imageseg_HZB} that the original image data is down-sampled to obtain cubic voxels with a side length of $\SI{20}{\nano\metre}$. 
Further down-sampling is performed to obtain images with cubic voxels the side length of which is equal to $\SI{40}{\nano\metre}$ and $\SI{80}{\nano\metre}$, respectively. 
For down-sampling by a factor of two, we remove every second slice along each of the three main spatial directions.
In order to obtain corresponding model realizations at these resolutions, all model parameters with physical units need to be converted to voxel units using the desired side length of a cubic voxel, see Table~\ref{tab:est_parameters} for the physical units of all model parameters. Dimensionless parameters such as $\gamma$ and $\mu$ remain unchanged at varying resolutions.

The descriptors considered here include the volume fraction of the solid phase and its specific surface area, \emph{i.e.}, the surface area per unit volume, which is estimated by the algorithm described in~\cite{ohser.2009}.
Moreover, we consider the  constrictivity and mean geodesic tortuosity that quantify the influence of bottleneck effects as well as the length of shortest transportation paths within a particular phase of the material, respectively.

The constrictivity, denoted by $\beta$, is  defined as the ratio $\beta=r_{\textup{min}}^2/r_{\textup{max}}^2$. Here, $r_{\textup{max}} = \mathrm{CPSD}^{-1}(0.5)$ is the maximum radius such that at least 50\% of the considered phase can be covered by spheres with radius $r_{\textup{max}}$ that are fully contained in the given phase, where $\operatorname{CPSD}\colon[0,\infty)\to[0,1]$ is the function defined in Eq.~\eqref{eq:CPSD_def}, where we replace $\Xi^{(c)}$ with $\Xi$ when the phase under consideration is the solid phase.
On the other hand, $r_{\textup{min}}$ is the maximum radius such that 50\% of the  phase under consideration can be reached by a ball with radius $r_{\textup{min}}$ intruding into this phase from a pre-defined side of the material, where $r_{\textup{min}}$ can be considered as the size of a typical bottleneck~\cite{n.2019}. It holds that $0\leq\beta\leq 1$, where the smaller $\beta$ is the stronger the bottleneck effects are. 

The mean geodesic tortuosity $\tau$ is the expected length (normalized by the material's thickness) of the shortest transportation path within the considered phase from a predefined starting point on one side of the material to the opposite side. 
By definition, it holds that $\tau\geq 1$, where values of $\tau$ closer to $1$ indicate shorter transportation paths within the given phase. 
Formal definitions of the quantities $\tau$, $r_{\textup{min}}$, $r_{\textup{max}}$ and $\beta$ and their respective estimators in the framework of stationary random closed sets can be found in~\cite{n.2018}. 

\begin{figure}[h]
    \centering
    \includegraphics[width=.9\textwidth]{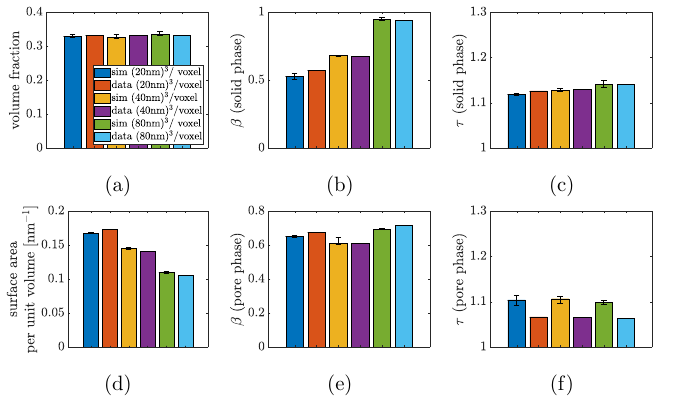}
    \caption{Comparison of morphological descriptors of  realizations of $\Xi$ and tomographic image data at varying voxel resolutions, where volume fraction~(a), specific surface area~(d), constrictivity $\beta$ of  solid~(b) and  pore phase~(e), as well as mean geodesic tortuosity $\tau$ of  solid~(c) and  pore phase~(f) are depicted. For model realizations, the bar heights show the average, while the error bars show the minimum and maximum obtained for  five realizations of $\Xi$.
    } 
    \label{fig:validation_betatau}
\end{figure}

Numerical values obtained for these descriptors for tomographic image data and model realizations
are visualized in Figure~\ref{fig:validation_betatau}. 
For a detailed discussion on these results, see also Section~\ref{sec:goodnessoffit}.

\begin{figure}[h]
    \centering
    \includegraphics[width=\textwidth]{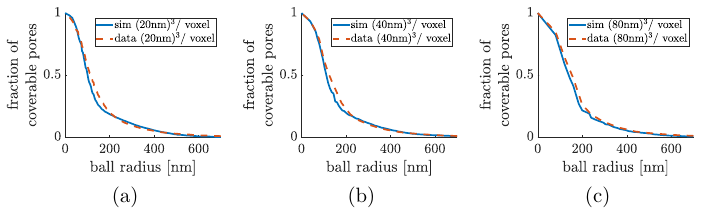}
    \caption{Comparison of continuous pore size distributions of  realizations of $\Xi$ and tomographic image data for cubic voxels of side lengths equal to \SI{20}{\nano\metre}~(a),  \SI{40}{\nano\metre}~(b), and \SI{80}{\nano\metre}~(c). }
    \label{fig:cpsd}
\end{figure}

Furthermore, the numerical results obtained for the continuous pore size distribution of $\Xi$ are visualized in Figure~\ref{fig:cpsd}. As mentioned in Section~\ref{sec:poissonthinning}, the rate parameter $\theta$ of the Boolean model $\Xi^{(3)}$ was chosen in such a way that the  continuous pore size distributions
of  realizations of $\Xi$ and tomographic image data match closely. Nevertheless, in Figure~\ref{fig:cpsd}, we consider this descriptor also for realizations of $\Xi$ at the coarser resolutions.

\begin{figure}[h]
    \centering
    \includegraphics[width=\textwidth]{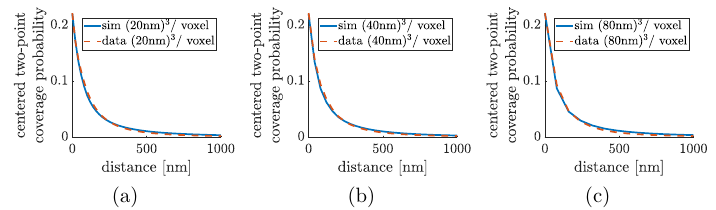}
    \caption{Comparison of centered two-point coverage probability functions computed for  realizations of $\Xi$ and tomographic image data, for cubic voxels of side lengths equal to \SI{20}{\nano\metre}~(a),  \SI{40}{\nano\metre}~(b), and \SI{80}{\nano\metre}~(c).   }
    \label{fig:2pc}
\end{figure}

Finally, we consider the values  $C(h)-V^2$ for $h\geq 0$ of the centered two-point coverage probability function of $\Xi$, where $C(h)$ is defined as in Eq.~\eqref{eq:twopointcovdef} by replacing $\Xi^{(2)}$ with $\Xi$, and $V$ is the volume fraction of $\Xi$. This function is compared to its empirical counterpart computed from tomographic image data, see Figure~\ref{fig:2pc}.

\subsection{Structure-property relationships}\label{sec:vary_graphite}
Since the combined stochastic 3D model $\Xi$ introduced in Section \ref{sec:model_desc}
can be characterized by a small number of
 interpretable parameters, it is possible to systematically modify these parameters in order to generate virtual but realistic nanostructures with varying effective properties. 
This opens the path for virtual materials testing~\cite{n.2019,prifling.2021c}, where effective properties of the simulated nanostructures can be analyzed by numerical modeling and simulation, in order to derive structuring recommendations prior to real experimental manufacturing of new materials. 

In particular, the stochastic 3D model $\Xi$ developed in the present paper allows us to separately modify the parameters of the stationary random closed sets $\Xi^{(1)}$, resembling the graphite particles, and $\Xi^{(2)}$, resembling the  mixture of carbon black and binder. In the following, we showcase the approach of virtual materials testing by adjusting the fraction of graphite particles in the modeling component $\Xi^{(1)}$, where we vary the intensity $\lambda_X$ of the underlying Poisson point process.
The values considered for $\lambda_X$ and the corresponding volume fractions $V_1$ of  $\Xi^{(1)}$ are given in Table~\ref{tab:lambda_x_values}. 
\begin{table}[h!]
    \centering
    \begin{tabular}{l|c}
        value of $\lambda_X$~[\unit{\per\nano\metre\cubed}]  &  volume fraction $V_1$ of $\Xi^{(1)}$ \Bstrut\\\hline \Tstrut
        $2.21 \cdot 10^{-11}$    &  0.0381 \\
        $3.55 \cdot 10^{-11}$    &  0.0605 \\
        $4.93 \cdot 10^{-11}$    &  0.0830  \\
        $6.36 \cdot 10^{-11}$    &  0.1057  \\
        $7.81 \cdot 10^{-11}$    &  0.1283 \\
        $9.33 \cdot 10^{-11}$    &  0.1512  \\
        $1.09 \cdot 10^{-10}$    &  0.1744
    \end{tabular}
    \caption{Values of the parameter $\lambda_X$ and corresponding volume fraction  $V_1$ of the stationary random closed set $\Xi^{(1)}$ used for the modeling of varying graphite levels.}
    \label{tab:lambda_x_values}
\end{table}

The virtual, yet realistic, image data of nanoporous binder-conductive additives with varying amounts of graphite particles can be used for predictive simulations of effective properties, \emph{i.e.}, based on these virtual nanostructures we can investigate structure-property
relationships. More precisely,  we can quantitatively study the influence of graphite particles on effective transport properties in the nanoporous binder-conductive additive phase, which have
a crucial impact on electrochemical processes in the cathode and thus on the performance of
battery cells.

To achieve this goal, we consider several morphological descriptors (volume fraction, constrictivity, mean geodesic tortuosity) as well as the so-called M-factor  for both,  the solid phase  and the pore space of $\Xi$. 
The M-factor, denoted by $M$ in the following,  quantifies the apparent decrease in diffusive transport resulting from obstacles along transport paths through porous media, \emph{i.e.}, 
it is the ratio of effective over intrinsic conductivity, when the solid phase is considered, and the ratio of effective over intrinsic diffusivity in the case of the pore space~\cite{holzer.2023}. 
Alternatively, it is the inverse of the so-called MacMullin number~\cite{Landesfeind_2016}.
Formally, the M-factor can be defined by 
\begin{equation}\label{for.tau.fac}
M=\frac{\varepsilon }{ \tau_{\mathrm{eff}}},
\end{equation}
where $\varepsilon$ is the volume fraction of the given phase (i.e. $\varepsilon = V$ or $\varepsilon = 1-V$ depending on whether we consider pore or solid phase) and $\tau_{\mathrm{eff}}$ is its effective tortuosity. The latter can be computed using the Matlab application TauFactor~\cite{copper.2016}.
To compute the effective tortuosity $\tau_{\mathrm{eff}}$ of the solid phase, the intrinsic conductivity of $\Xi^{(1)}$, resembling the graphite particles, was set to be $100$ times larger than the intrinsic conductivity of $\Xi^{(2)}$, resembling the carbon black and binder phase, according to Table~2 in~\cite{marinho.2012}. We therefore assume a normalized dimensionless conductivity of $1$ for the graphite particles and dimensionless conductivity of $0.01$ for the carbon black and binder phase. 
Note that $\tau_{\mathrm{eff}}$ is not a purely morphological descriptor, but it is determined by solving certain differential equations numerically~\cite{copper.2016}, which can be computationally expensive. Alternatively, an approximation $\widehat M$ of $M$ can be obtained by using regression formulas like

\begin{equation}\label{ana.reg.for}
    \widehat{M} = \frac{\varepsilon^{a}\beta^{b}}{\tau^{c}},
\end{equation}
where $\beta,\tau$ denote the constrictivity and the mean geodesic tortuosity of the phase under consideration, and $a,b,c\geq 0$ are some constants. This kind of regression is of the same form as those considered in~\cite{neumann.2023,n.2019}, except with adapted coefficients. Namely, in our case it turned out that $a=2.1939$, $b=0$ and $c=5.0152$ leads to the best fit. 
Recall that for computing the M-factor of the solid phase, separate intrinsic conductivities for voxels belonging to graphite particles and the carbon black and binder phase, respectively, are used, where the intrinsic conductivity of graphite is assumed to be 100 times higher than that of the carbon black and binder phase. 
Therefore, we cannot expect to obtain the same coefficients as in~\cite{neumann.2023,n.2019} since these previous studies consider phases with constant normalized intrinsic conductivity of $1$. The benefit of the approximation $\widehat M$ of $M$ given in Eq.~\eqref{ana.reg.for} is that it only depends on the morphological descriptors $\varepsilon,\tau,\beta$ which can be determined directly from voxelized image data, \emph{i.e.}, without solving differential equations, which is typically more efficient than the computation of the effective tortuosity $\tau_{\mathrm{eff}}$.

Furthermore, we consider a linear fit of the form 
\begin{equation}\label{eq:linear.M.fit}
    \widetilde{M} = c_1 \cdot V_1 + c_0,
\end{equation}
where $V_1$ is the volume fraction of $\Xi^{(1)}$ and $c_1,c_0\in\R$ are some constants. This yields a linear approximation of the M-factor $M$ in dependence of the volume fraction of graphite particles $V_1$. We stress that such a linear fit can only be reasonable within the considered range of values for $V_1$ and can not be extrapolated arbitrarily. Nevertheless, Eq.~\eqref{eq:linear.M.fit} is useful in demonstrating the near-linear dependence of $M$ on $V_1$ within the considered range of values for $V_1$. The values of $c_0$ and $c_1$ were determined by least square estimation, which resulted in $c_0=-0.02061$ and $c_1 = 0.5588$ if the considered phase is the solid phase, and $c_0=0.4556$ and $c_1 = - 1.874$ if the considered phase is the pore phase.

\begin{figure}[h]
    \centering
    \includegraphics[width=0.9\textwidth]{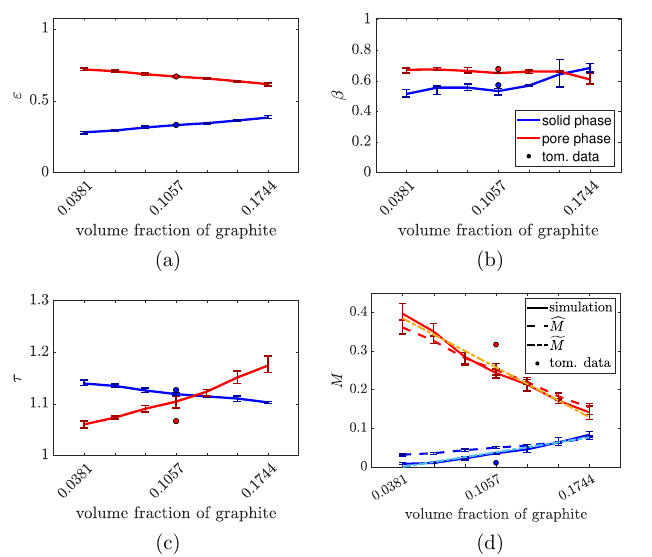}
    \caption{Comparison of morphological descriptors of solid phase (blue) and pore phase (red) for model realizations of $\Xi$ with varying volume fractions of graphite particles. Depicted are the volume fraction $\varepsilon$ (a),  constrictivity $\beta$ (b), and mean geodesic tortuosity $\tau$ (c) as well as the M-factor $M$~(d). Panel (d) also shows the approximations $\widehat{M}$ and $\widetilde{M}$. The (solid and dashed) lines  show the averages obtained for the five realizations of $\Xi$, while the error bars indicate the corresponding minimum and maximum. In all panels, the descriptors obtained from tomographic image data are shown as colored disks.}
    \label{fig:graphite_betatau}
\end{figure}

The results which have been  obtained for the volume fraction $\varepsilon$, constrictivity $\beta$, and mean geodesic tortuosity $\tau$ of the solid phase and the pore space of $\Xi$, when varying the volume fraction of graphite particles as stated in  Table~\ref{tab:lambda_x_values}, are visualized in Figures~\ref{fig:graphite_betatau}a--c, where  averages over five different realizations of $\Xi$ have been used. 
The corresponding results obtained
for $M$, $\widehat M$ and $\widetilde{M}$ by means of Eqs.~\eqref{for.tau.fac}--\eqref{eq:linear.M.fit} are shown in Figure~\ref{fig:graphite_betatau}d.

\section{Discussion}\label{sec:discussion}
\subsection{Goodness of model fit}\label{sec:goodnessoffit} For both, tomographic image data and model realizations, various morphological descriptors of solid and pore phase have been computed at different voxel resolutions, see Figure~\ref{fig:validation_betatau}.
The fact that the volume fractions of tomographic image data and model realizations are nearly identical for all  resolutions is an expected result, as the model was directly calibrated based on the volume fraction estimated from 3D image data and the point-count method yields and unbiased estimator of the volume fraction, independently of the underlying voxel resolution, see Section 6.4.2 of~\cite{chiu.2013}. The surface area per unit volume of   $\Xi$ has also been used for model calibration via the intrinsic volumes of $\Xi^{(1)}$, and  indirectly also via the covariance function of $\Xi^{(2)}$, as the surface area of a Gaussian excursion set can, under mild assumptions, be expressed in terms of the volume fraction of $\Xi^{(2)}$ and the covariance function of the underlying Gaussian random field, see Eqs.~(6.163)--(6.165) of~\cite{chiu.2013}. However, the influence of $\Xi^{(3)}$ on the surface area of $\Xi$ is a priori unclear. Nevertheless, model realizations and tomographic image data agree well with respect to surface area per unit volume of $\Xi$ across all three resolutions. 
The continuous pore size distribution estimated from tomographic image data, shown in Figure~\ref{fig:cpsd}, enters model calibration only at the resolution of $\SI{20}{\nano\metre}$. On coarser resolutions, the continuous pore size distribution of $\Xi$ remains nearly unchanged for both model realizations and image data.

The remaining morphological descriptors considered in Section~\ref{sec:modelval} for model validation were not used for model calibration.
While constrictivity is a descriptor that is rather sensitive to small changes of the nanostructure, it is still accurately reproduced at all three resolutions for both solid and pore phase. 
The mean geodesic tortuosities of model realizations and tomographic image data are in close alignment for the solid phase, but show a moderate deviation for the pore phase. 
This is likely due to 
the simplifying  model assumptions made in Section~\ref{sec:model_desc}.
For example, in the tomographic image data one can observe that larger pores often accumulate close to larger graphite particles, which is not captured by the model as it would drastically increase its complexity and, thus, the number of model parameters.
Finally, the centered two point coverage functions of model realizations at different resolutions are nearly identical to those of the tomographic image data, see Figure~\ref{fig:2pc}. Note that while we used the two point coverage function in order to calibrate the parameters of $\Xi^{(2)}$, it is a priori unclear whether the two point coverage function of the whole model $\Xi$ would agree with that of the complete tomographic image data. We considered multiple descriptors, each of which captures different morphological properties of the nanostructures. Overall, the descriptors considered in this paper show a good agreement between image data and model realizations at all three resolutions and appropriately validate the combined stochastic 3D model $\Xi$. In particular, one should keep in mind that $\Xi$ is characterized by only eight model parameters, but is still able to reproduce the complex nanostructure of the binder-conductive additive phase to a large extent.

\subsection{Effects of varying resolutions}
One of the goals of developing this model is to generate input data for subsequent numerical simulations, which are often limited by the size of data that they can process and therefore need to operate at coarser resolutions.
In order to assess the goodness of  model fit at varying resolutions, the tomographic image data was down-sampled by removing every second slice along each of the main spatial directions, while the model parameters have been converted to voxel units of the desired voxel size according to their physical units shown in Table~\ref{tab:est_parameters} to generate samples of the corresponding resolution. 
We then compare morphological descriptors of tomographic image data and model realization at these coarser resolutions. Some of the descriptors which are considered in this paper, in particular specific surface area and constrictivity, show significant differences for the varying resolutions.
This is not surprising, as details of  rough surfaces and fluctuations in pore widths are increasingly lost on lower resolutions, which decreases the  surface area and affects the overall influence of bottlenecks, quantified by  constrictivity. This effect is intrinsic to the nature of voxelized image data. Most importantly, however, the trend is accurately reproduced by the stochastic 3D model $\Xi$, such that the accordance of all descriptors considered in this paper  is at  a similar level for the three different resolutions.

\subsection{Virtual materials testing} One of the strengths of  parametric models is their ability to directly control the resulting realizations through interpretable parameters, which opens up the path for virtual materials testing by generating virtual but realistic 3D morphologies for varying parameters and analyzing their properties.
We illustrate this approach using the stochastic 3D  model $\Xi$ introduced in this paper as an example. Namely, by adjusting the parameters of the Boolean model $\Xi^{(1)}$, we are able to generate virtual nanostructures of the additive matrix with varying volume fractions of graphite, for which various morphological descriptors are computed, see Figure~\ref{fig:graphite_betatau}.

Clearly,  as shown  in Figure~\ref{fig:graphite_betatau}, changes in the volume fraction of graphite particles result in  corresponding changes of the volume fraction of the entire solid phase. The constrictivity  fluctuates only slightly across the considered volume fractions of graphite particles, which indicates that there might be no clear connection between the amount of graphite particles and the constrictivity of the binder-conductive additive phase.
However, further investigations are necessary in order to understand this behavior. On the other hand, the mean geodesic tortuosity as well as the M-factor show  clear trends. Here, a larger amount of graphite particles leads to an increase in tortuosity of the pore space and a decrease in tortuosity of the solid phase.
This is a reasonable effect, as the elongated shapes of  graphite particles serve to create new transport paths in the solid phase, thereby decreasing the tortuosity of the solid phase, while blocking some paths in the pore space, thereby increasing the tortuosity of the pore space. 
It can be seen that the M-factor follows a roughly linear trend in dependence of the volume fraction of graphite particles, as indicated by the linear fit $\widetilde{M}$ shown in Figure~\ref{fig:graphite_betatau}d.

The computation of the M-factor on the solid phase of tomographic image data yields a value of $0.0112$ which is roughly a third of the mean value of $0.0357$ computed for the solid phase of model realizations that correspond to the $10.57\%$ graphite observed in tomographic image data. 
On the other hand, the values of the M-factor obtained for the pore phase of tomographic image data and model realizations agree much better to each other, with a value of $0.3165$ for tomographic image data compared to  $0.2429$ for model realizations.

We believe that this disagreement of values obtained for the M-factor is due to simplifications in the model. For the pore phase, it is a result of the deviations in mean geodesic tortuosities already observed in Figure~\ref{fig:validation_betatau}f, which propagates to the fit of the M-factor, as predicted by Eq.~\eqref{ana.reg.for}. For the M-factor of the solid phase, the spatial arrangement of graphite particles plays a crucial role, as their conductivity is assumed to be 100 times larger than that of the carbon black and binder mixture. However, in the model $\Xi$ the random sets $\Xi^{(1)}$ and $\Xi^{(2)}$ are independent, which disregards spatial correlations between the graphite particles and the carbon black and binder mixture. This aspect still requires further investigation and can be improved in future work.

Finally, we  showed that it is possible to predict the numerically computed M-factor by means of a simple analytical regression that only depends on purely morphological descriptors which can easily be computed from voxelized image data. This further improves the efficiency of the virtual materials testing approach based on stochastic 3D modeling.

\section{Conclusion}\label{sec:conclusion}
We developed a stochastic 3D model for the nanoporous binder-conductive additive phase in battery cathodes. 
The model is calibrated based on tomographic image data and validated by comparing  various morphological descriptors of measured and simulated image data, which are not used for calibration. 
Moreover, model validation is performed at different resolutions of image data, which shows that realistic model realizations at different resolutions can be obtained by  scaling of  model parameters. 
The model accounts for the 3D morphology of carbon black covered with binder and, separately, of graphite particles by means of interpretable parameters. This allows for generating virtual but realistic morphologies of a binder-conductive additive phase with varying mixing ratios. 
In particular, it allows for virtual materials testing, where effective properties of  simulated nanostructures can be analyzed, in order to derive structuring recommendations prior to
real experimental manufacturing of new electrode materials.

\section*{Availability of data and materials}
The code for simulating the stochastic 3D model is publicly available as an R-package and can be downloaded using the link \texttt{https://CRAN.R-project.org/package=CBAModel}. The corresponding DOI is \texttt{10.32614/CRAN.package.CBAModel}. Data considered in this paper will be made available upon reasonable request.

\section*{Competing interests} The authors have no conflicts of interest to declare that are relevant to the content of this article.

\section*{Author contributions} Materials were synthesized by N. Bohn and J. R. Binder. 3D imaging and segmentation of gray scale images was performed by M. Osenberg, A. Hilger and I. Manke. Segmentation of graphite particles, stochastic 3D modeling, model calibration and the quantification of structure-property relationship were done by P. Gräfensteiner, V. Schmidt, and M. Neumann. All authors contributed in interpreting the obtained results.

\section*{Funding and acknowledgements}
This research is funded by the German Federal Ministry for Economic Affairs and Climate Action (BMWK) and granted through Project Management J\"ulich (03ETE039H, 03ETE039G, 03ETE039J). Publication is supported by TU Graz Open Access Publishing Fund.

\clearpage
\appendix
\section*{Appendix}
\setcounter{figure}{0}
\renewcommand\thefigure{A\arabic{figure}}

\begin{figure}[h]
    \centering
    \includegraphics[width=0.85\textwidth]{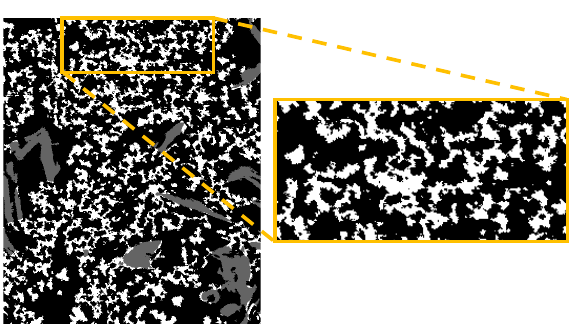}
    \caption{Visualization of the image data used for calibration of the excursion set $\Xi^{(2)}$. The chosen cutout shows a representative distribution of carbon black and binder without graphite particles or large pores. Carbon black and binder are represented in white, while the pore space is represented in black. The size of this cutout is $278\times 136 \times 124$ voxels.}
    \label{fig:app_gaussian_cutout}
\end{figure}

\begin{figure}[h]
    \centering
    \includegraphics[width=0.85\textwidth]{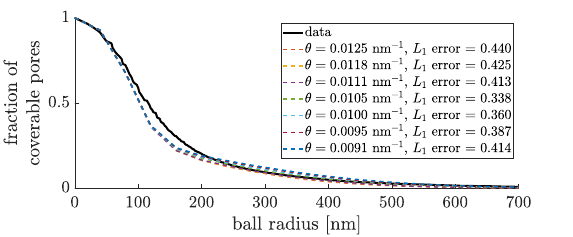}
    \caption{Comparison of continuous pore size distributions computed from tomographic image data (solid line) and from model realizations with differently chosen values of the rate parameter $\theta$ of $\Xi^{(3)}$ (dashed lines), where the dashed lines show the mean continuous pore size distributions taken over 3 model realizations for each value of $\theta$.   The results visualized in this figure have been obtained for a resolution of $\SI{20}{\nano\metre}$.}
    \label{fig:app_nu_choice}
\end{figure}

We provide some additional information regarding the calibration of the stochastic 3D model $\Xi$ introduced in this paper. Figure~\ref{fig:app_gaussian_cutout} shows the cutout of  tomographic image data used to estimate the parameters of the Gaussian excursion set $\Xi^{(2)}$. Figure~\ref{fig:app_nu_choice} shows a comparison of the continuous pore size distribution computed from tomographic image data with continuous pore size distributions of model realizations for different values of the rate parameter $\theta$ of $\Xi^{(3)}$.

\end{document}